\documentclass{PoS}

\title{Local Polyakov loop domains and their fractality}

\ShortTitle{Local Polyakov loop domains and their fractality}

\author{\speaker{Hans-Peter Schadler}\\
        Institut f\"ur Physik, Karl-Franzens Universit\"at Graz, 8010 Graz, Austria\\
        E-mail: \email{hps@abyle.org}}

\author{Gergely Endr\H{o}di\\
        Institut f\"ur Theoretische Physik, Universit\"at Regensburg, 94040 Regensburg, Germany\\
        E-mail: \email{gergely.endrodi@physik.uni-regensburg.de}}

\author{Christof Gattringer\\
        Institut f\"ur Physik, Karl-Franzens Universit\"at Graz, 8010 Graz, Austria\\
        E-mail: \email{christof.gattringer@uni-graz.at}}

\abstract{We discuss properties of local Polyakov loops in the deconfinement transition of SU(3) lattice gauge theory at finite temperature using the fixed scale approach. In particular we study spatial 
clusters where local Polyakov loops have phases near the same center elements of the gauge group. We present results for various properties of the center clusters, e.g., their percolation probability 
or their fractality and discuss the physical implications for temperatures below and above the phase transition.}

\FullConference{31st International Symposium on Lattice Field Theory - LATTICE 2013\\
		July 29 - August 3, 2013\\
		Mainz, Germany}

\begin{document}

\section{Introduction}
In recent years various experiments like the Large Hadron Collider at CERN or the 
Relativistic Heavy Ion Collider in Brookhaven, USA, have started to collect data and to analyze 
properties of the quark gluon plasma (QGP) such as its viscosity or opacity. Understanding the 
high temperature phase of matter is of fundamental importance for our deeper understanding of 
quantum chromodynamics (QCD). So far many properties of the QGP have been measured but for 
many of them we still lack a concluding theoretical explanation.

In recent years there have been attempts to describe phenomenological properties of the QGP 
\cite{centermodels} by using the old idea of center domains \cite{centerdomains}. These center domains 
are related to the static quark potential as they are defined by the phases of the Polyakov loop. Used 
in the description of the QGP they may be able to explain some of the phenomena observed in 
heavy ion collisions.

On the lattice one has direct access to the characteristic gauge configurations which allows one to directly 
study the center domains defined by the local Polyakov loop. Various such studies can be found in the literature
\cite{fortunato, OldStudies}, but so far  the temperature $T=(a(\beta)\,N_t)^{-1}$ was always changed by varying
the lattice constant $a(\beta)$, i.e., by changing the inverse gauge coupling $\beta$.
This has the major disadvantage that various effects related to temperature, the volume and the lattice spacing $a$
were mixed. For a clean study of the properties of the center clusters one would like to disentangle these effects. 
This can be accomplished by using the fixed scale approach, i.e., one works at a fixed inverse gauge coupling $\beta$ 
(and thus fixed lattice constant $a$),  and drives the temperature by varying the temporal lattice extent $N_t$.

In this preliminary study we analyze properties of the center domains of pure SU(3) lattice gauge theory 
using the Wilson plaquette action for $N_s^3 \times N_t$  lattices of spatial sizes $N_s=30,\;40$ and $N_s=48$. 
The temperature is driven by varying the inverse temporal lattice extent $N_t=2,...,20$ and we perform calculations 
for three different lattice spacings $a = 0.1117,\;0.0677$ and $0.0481$ fm. 

\section{General properties of the Polyakov loop}
In the continuum the local Polyakov loop at a spatial point $x$ is defined as the trace of the path-ordered exponential 
\begin{equation}
	L(x) \; = \; \mbox{Tr} \, {\cal P}  \exp \bigg( \int_0^{1/T} \! A_4(x,t) \, dt \bigg) \;,
\end{equation}
where $t$ is the euclidean time which is integrated between 0 and the inverse temperature $1/T$ (we set the Boltzmann constant to $k_B=1$). 
On the lattice, the local Polyakov loop is given by the trace of the product over all time-like links $U_4(x,t)$
\begin{equation}
	L({x}) \; = \; \mbox{Tr} \, \prod_{t=1}^{N_t} U_4(x,t) \; ,
\end{equation}
which is obviously a gauge invariant object. Under a center transformation, i.e., multiplying 
all temporal links $U_4(x,t_0)$ on a certain time-slice $t_0$ with a center element $z \in \{1, e^{i2\pi/3},e^{-i2\pi/3} \}$ 
of SU(3), the local Polyakov loop transforms non-trivially and picks up this center element as a factor, 
$L(x) \rightarrow zL(x)$. Thus it serves as an order parameter for the breaking of the center symmetry 
which for pure gauge theory is a symmetry of the action. Above the critical temperature $T>T_c$ the center 
symmetry is spontaneously broken, manifest in a non-vanishing expectation value for the local Polyakov loop 
$\langle L(x) \rangle \neq 0$. In the case of dynamical fermions, the fermion determinant breaks the center 
symmetry explicitly and the system favors the trivial center sector around $z=1$ resulting in a non-vanishing 
Polyakov loop expectation value already below the transition.

From previous work \cite{OldStudies} it is known that the phase $\theta(x)$ of the Polyakov loop plays a special role (we write 
$L(x) = |L(x)| \exp(i\theta(x))$). Below the phase transition the phases $\theta(x)$ are equally distributed in all center sectors with peaks 
of equal hight at $\theta(x)=0,\pm 2\pi/3$ corresponding to the center elements of SU(3). Averaging over the phases leads to the 
vanishing expectation value $\langle L(x) \rangle = 0$. Above the phase transition, the behavior is different. The system will spontaneously 
choose a preferred center sector and the distribution of the phases $\theta(x)$ will show a peak at the corresponding center element 
which leads to a non-zero expectation value $\langle L(x) \rangle$.


\begin{figure}[b]
	\centering
	\includegraphics[width=0.5\textwidth,clip]{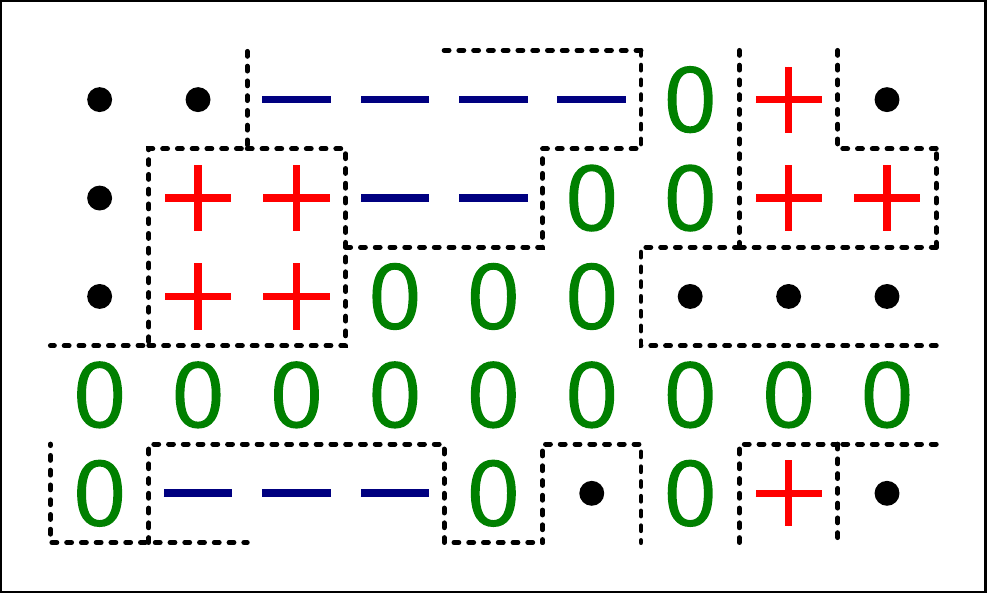}
	\caption{A possible cluster configuration with clusters in all three sectors $0,\pm 1$. The black dots correspond to spatial points which are removed for some 
	$f>0$. Some of the clusters are connected via the periodic boundary conditions and the sector $0$ (green) percolates, i.e., 
       it extends over the whole lattice in all directions.}
	\label{DrawingCluster}
\end{figure}

\section{Definition of center sectors and clusters}
Based on the qualitative discussion of the properties of the local Polyakov loops $L(x)$ and its phases $\theta(x)$, 
we assign to every spatial point $x$ the center sector number $n({x})$ according to the following prescription,
\begin{equation}
	n({x})  = \left\{ \begin{array}{rl}
	-1 & \; \mbox{for} \;\; \theta(x) \, \in \, 
	[\,-\pi + \delta \; , \; -\pi/3 - \delta \, ]\\
	0 & \; \mbox{for} \;\; \theta(x) \, \in \, 
	[\,-\pi/3 + \delta \, , \, \pi/3 - \delta \, ] \; , \\
	+1 & \; \mbox{for} \;\; \theta(x) \, \in \, 
	[\,\pi/3 + \delta \, , \, \pi - \delta \,]
	\end{array} \right.
\end{equation}
with a real and non-negative parameter $f$,
\begin{equation}
	\delta \; = \; f \, \frac{\pi}{3} \; \; , \;\; \;\; f \in [0,1] \; ,
\end{equation}
which we refer to as the cut parameter. The role of the parameter is to allow for removing sites $x$ from the cluster analysis where the local Polyakov 
loops $L(x)$ do not clearly lean towards one of the center elements: For $f=0$ no sites are removed, while for $f\rightarrow1$ all sites are removed,
and values in between the two extrema allow a gradual removal of "undecided" sites. Fig.~\ref{DrawingCluster} illustrates a possible decomposition of a
(2-dimensional) lattice into sites with $n(x) = 0, \pm 1$ and sites that are removed with a finite value of $f$.

\begin{figure}[t]
	\centering
	\includegraphics[width=0.32\textwidth,clip]{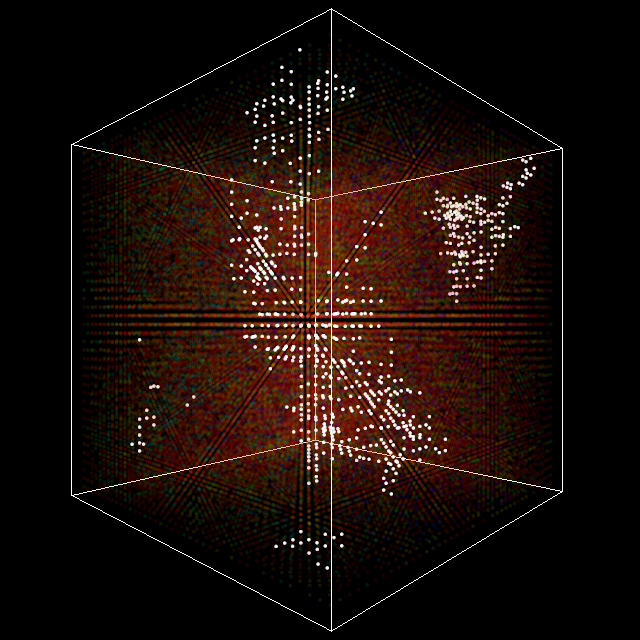}
	\includegraphics[width=0.32\textwidth,clip]{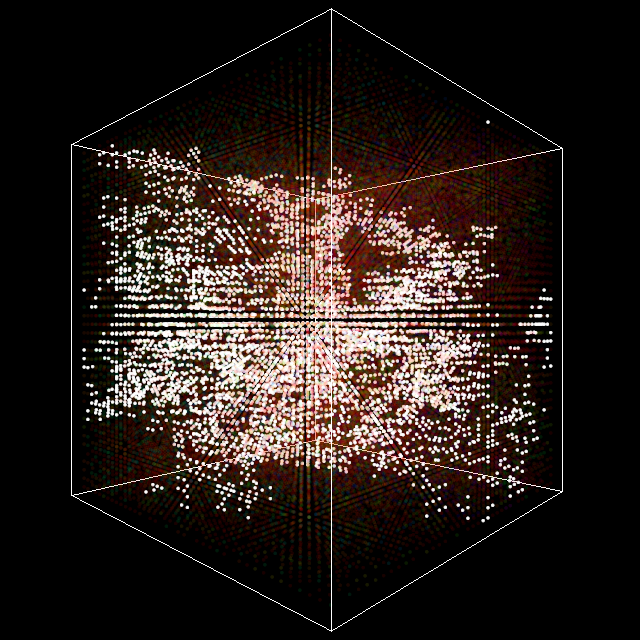}
	\includegraphics[width=0.32\textwidth,clip]{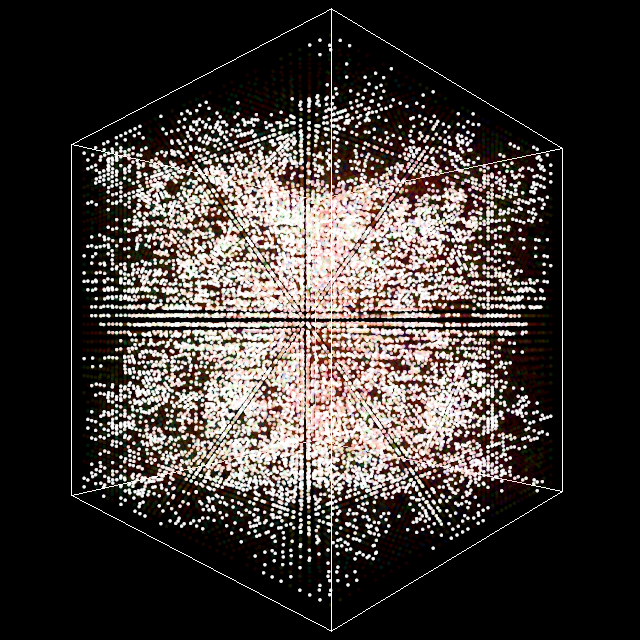}
	\caption{The largest center cluster on a $40^3\times N_t$ lattice for 
	temperatures $T/T_c=0.90,\,0.99,\,1.10$ (from left to right) at a cut parameter $f = 0.3$.}
	\label{Cluster3D}
\end{figure}

Center clusters are now defined in the following way: Two neighboring spatial points 
$x$ and $y=x \pm \hat\mu$ belong to the same cluster if $n({x})=n({y})$, i.e., if they are 
near the same center element. In Fig.~\ref{DrawingCluster} 
we show a possible schematic configuration on a 2-D slice of the lattice and in Fig.~\ref{Cluster3D} 
we show the largest cluster on a $N_s=40$ lattice at three different temperatures for a cut parameter of $f=0.3$. 
It is obvious that when the temperature increases above $T_c$, the largest cluster dramatically increases in size and starts to percolate. 
Exactly these changes of the center clusters near $T_c$ are at the focus of this study. 

\begin{figure}[b]
	\centering
	\includegraphics[width=0.5\textwidth,clip]{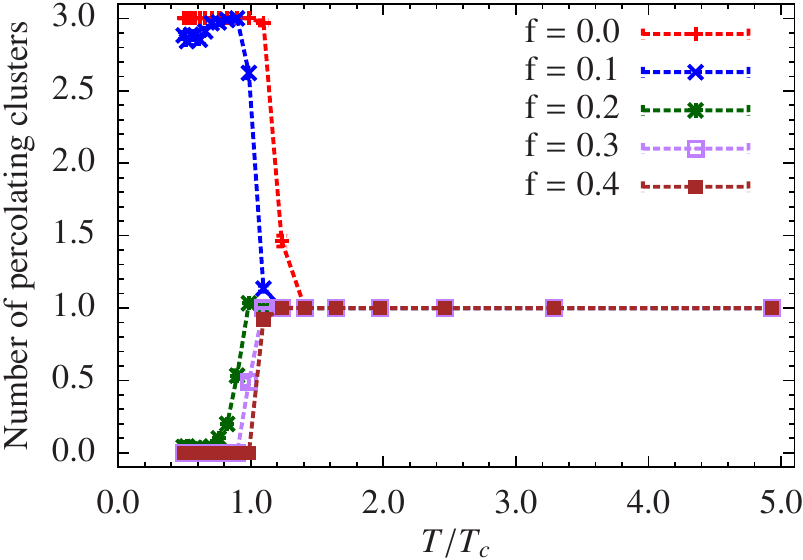}
	\caption{Number of percolating clusters as a function of the temperature for different cut parameters $f$. 
	We show the results for the $N_s=48$, $a=0.0677$ fm ensemble.}
	\label{perc}
\end{figure}

Before we discuss properties of the clusters, we need to illuminate a 
little bit the role of the cut parameter $f$. We start with an excursion to percolation 
theory and recall some fundamental properties: In random percolation theory the threshold 
for site percolation on a simple cubic 3-D lattice is $p = p_c=0.3116$ \cite{PercolationProbability}, 
where $p$ is the occupation probability for a lattice site, i.e., if the occupation probability $p$ is 
larger than $p_c$ we expect to find a percolating cluster. If we assume random distribution of the 
values $n(x)\in\{0,\pm 1\}$ and use a cut parameter $f=0$ (no sites removed), we get for every 
center sector an occupation probability of $p=1/3>0.3116 = p_c$, and thus expect that without cut 
there is a percolating cluster in each of the three center sectors. This is confirmed in Fig.~\ref{perc} 
where we show the number of percolating clusters on a $N_s=48$ lattice as a function of the temperature 
$T$ for different values of the cut parameter $f$. We observe the following: For $f=0$ we indeed find three 
percolating clusters for $T<T_c$, one for every center sector. For temperatures
$T>T_c$ one of the center sectors will become dominant and the number of percolating clusters drops to 
one.  For sufficiently large $f$ the picture is reversed, and below $T_c$ there is no percolating 
cluster while for $T > T_c$ a single percolating cluster emerges (compare Fig.~\ref{Cluster3D}).

\section{Cluster weight}

\begin{figure}[b]
	\centering
	\includegraphics[width=0.49\textwidth,clip]{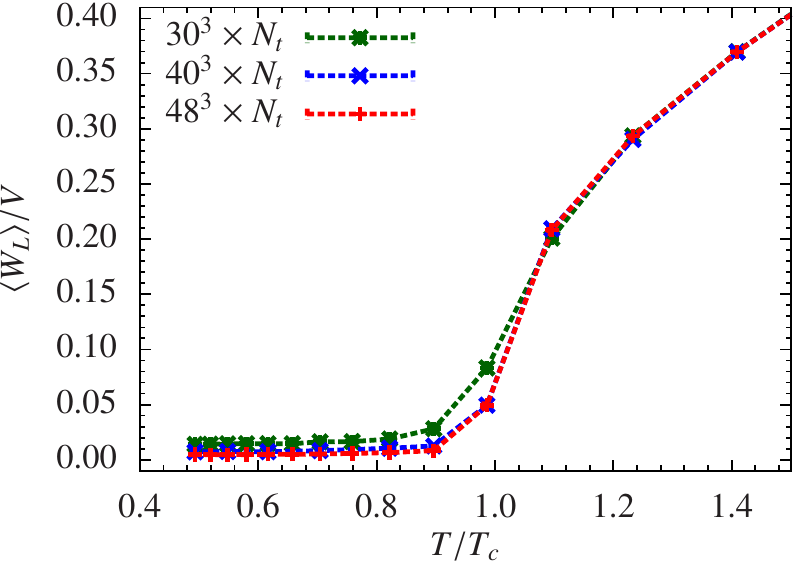}
	\includegraphics[width=0.49\textwidth,clip]{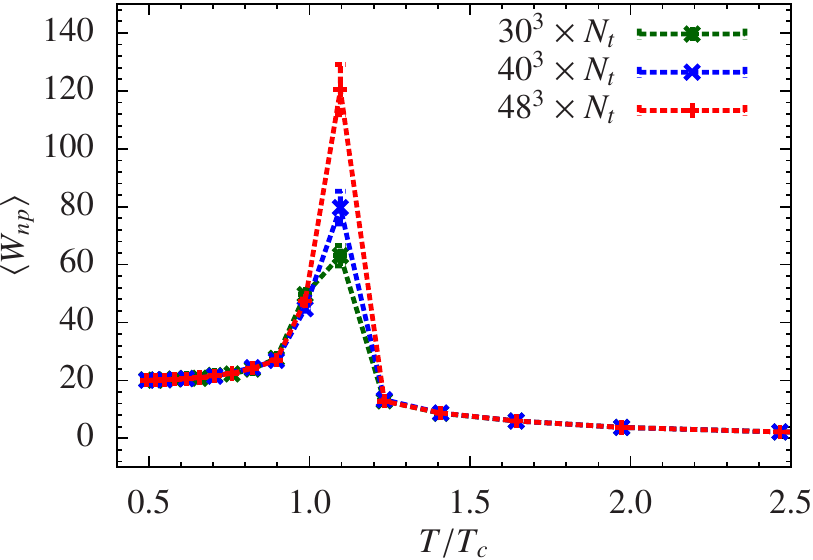}
	\caption{Comparison of the weight of the largest cluster (l.h.s.) and of the average non-percolating clusters (r.h.s.) for three different lattice volumes as a function of the temperature $T$. We show the results for the $a=0.0677$ fm ensemble with a cut parameter $f=0.3$.}
	\label{Weight}
\end{figure}

Simple to calculate, but nevertheless important, are observables related to the 
number of sites which belong to a cluster, i.e., the cluster weight. We study 
two different definitions: In Fig.~\ref{Weight} (l.h.s.) we show the average weight 
of the largest cluster, i.e., the number of lattice sites which belong to this cluster, 
normalized by the 3-volume $V=N_s^3$ for three different spatial volumes. 
The cut parameter was chosen to be $f=0.3$ and we work at $a=0.0677$ fm. 
Below the critical temperature the cluster weight does not depend on the temperature 
but shows a plateau. Furthermore, these plateaus have different values for the three 
different volumes, which indicates that the cluster weight is only weakly dependent on the 
volume below $T_c$ and the visible volume dependence in Fig.~\ref{Weight} comes solely 
from the normalization. At the critical temperature the weight of the largest cluster starts 
to increase as the cluster starts to percolate. From the different volumes we observe the 
usual rounding which is related to finite volume effects. Above the critical temperature 
$T>T_c$ there is perfect agreement between the different volumes and for high $T$ 
the weight tends to $\langle W_L \rangle/V=1$.

Also an interesting quantity, which serves as a susceptibility-like observable, is the average weight $W_{np}$ 
of the non-percolating clusters. We define it as (see, e.g., \cite{staufer, fortunato})
\begin{equation}
	w_s = \frac{n_s s}{\sum_{s^\prime} n_{s^\prime} {s^\prime}} \qquad , \qquad
	W_{np} \; = \; \sum_s w_s \, s \; = \; \frac{\sum_s n_s \, s^2}{\sum_{s^\prime} n_{s^\prime} \,  {s^\prime}} \; ,
	\label{defeq}
\end{equation}
where $s$ is the size of the cluster and $n_s$ is the number of clusters of size $s$. 
With this definition, $w_s$ is the probability that an occupied site belongs to a cluster of size $s$. 
Using this, we define the average weight $W_{np}$ of the non-percolating clusters as
given in the second equality of (\ref{defeq})
where all sums run over all non-percolating clusters.

We study the expectation value $\langle W_{np}\rangle$ as a function of the temperature, 
again for three different volumes (Fig.~\ref{Weight}, r.h.s.), and find a peak at the critical 
temperature $T=T_c$. Furthermore, also the expected scaling of the peak height with the 
lattice volume can be observed.

\section{Fractal dimension of the center clusters}

\begin{figure}[b]
	\centering
	\includegraphics[width=0.49\textwidth,clip]{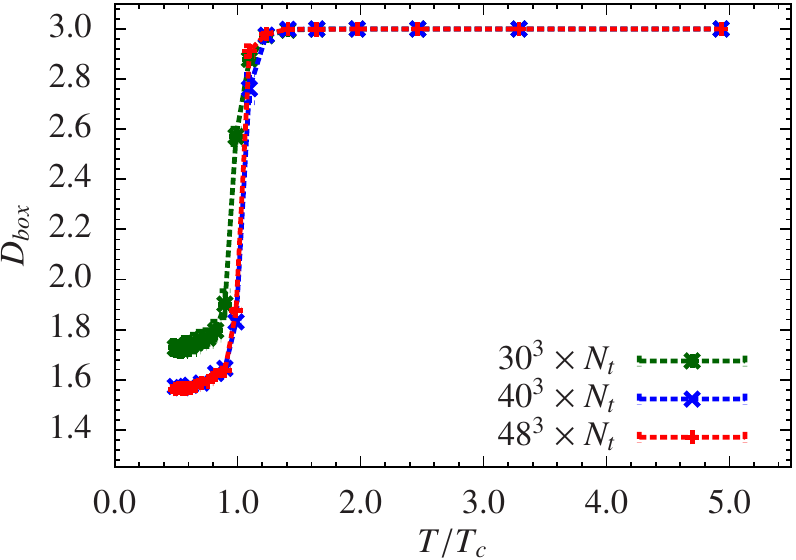}
	\caption{Box counting dimension $D_{box}$ as a function of the temperature for different spatial volumes. We use our $a=0.0677$ fm ensemble. The cut parameter $f$ is tuned such that the physical radius of the largest cluster at the lowest temperature is $R_{phys}=0.5$ fm. For $N_s=40$ we observe the usual rounding effect around $T_c$, while the $N_s=30$ results are plagued by more severe finite size effects which affect the quality of the fits and the resulting fractal dimension below $T_c$.}
	\label{FractDim}
\end{figure}

The last property that we want to discuss is the fractal dimension of the largest cluster. We determine the fractal dimension via the box counting method. In this method, one counts the number of boxes $N(s)$ of size $s$ needed to cover the whole cluster. In the limit of small $s$ we expect to find a behavior given by
\begin{equation}\label{eq:box}
	N(s) = C\,s^{-D_{box}} \; .
\end{equation}

In Fig.~\ref{FractDim} we plot the box counting dimension, obtained by a fit of the data to Eq.~(\ref{eq:box}), for our $a=0.0677$ fm ensemble as a function of the 
temperature for three different spatial volumes. The fractal dimension changes drastically with the temperature: While below $T_c$ the dimension is around $D_{box} 
\approx 1.6$, it shows a drastic increase at the critical temperature and approaches $D_{box} = 3.0$ already slightly above $T_c$. The $N_s=30$ ensemble shows 
strong finite size effects below $T_c$ which makes a reliable determination of $D_{box}$ challenging. For the $N_s=40$ ensemble, we only observe a mild rounding at 
$T=T_c$.

\section{Conclusion and final remarks}
We start our concluding words with a comparison to percolation theory. The analog of the 
temperature in percolation theory is the occupation probability $p$. Due to the dominance 
of a single percolating cluster above $T_c$, the occupation probability for the corresponding 
sector becomes larger with increasing temperature and at some point reaches the critical value 
$p_c$ where a cluster of the dominant sector starts to percolate. The change of the temperature 
therefore corresponds to changing the occupation probability in favor of one sector which is spontaneously chosen 
by the system in the spontaneous breaking of the center symmetry. 
This behavior can also be observed if one sets the cut parameter to $f=0$. The corresponding 
transition would then be from three equally distributed sectors with $p_i>p_c,\, i=1,2,3$, i.e., each 
with a percolating cluster, down to only one dominating sector with, for example, $p_{1,2}<p_c$ and $p_3 \geq p_c$
(compare Fig.~\ref{perc}).

The analysis of the weight and the fractal dimension of the largest cluster show that the clusters are small and 
highly fractal objects with $D_{box}\approx 1.6$ below the critical temperature $T<T_c$. At the critical temperature
the fractal dimension quickly jumps to $D_{box}=3.0$, while the weight of the largest cluster starts to increase steadily towards 
$\langle W_L \rangle/V=1$.

We remark that the cut parameter $f$ can be related to a physical quantity, 
the radius of the largest cluster at $T\ll T_c$, which changes as a function of $f$. The radius of 
the largest cluster in physical units for some temperature
$T < T_c$ can then be adjusted to some fixed value, e.g., to 0.5 fm, by adjusting $f$. 
Thus by suitably adjusting $f$ we can set the scale and use this procedure to compare 
results from lattices with different lattice constants $a$. This procedure and  other properties of the center clusters will be 
discussed in detail in our upcoming publication \cite{Paperinprep}.

\section*{Acknowledgments}
We thank S.\ Bors\'anyi, J.\ Danzer, M.\ Dirnberger, A.\ Maas, B. M\"uller and A. Sch\"afer for interesting discussions. In addition, the speaker wants to thank  T.\ Kloiber, C.\ Lang and A.\ Schmidt. This work is partly supported by DFG TR55, ``{\sl Hadron Properties from Lattice QCD}'' and by the Austrian Science Fund FWF Grant.\ Nr.\ I 1452-N27. H.-P.~Schadler is funded by the FWF DK W1203 ``{\sl Hadrons in Vacuum, Nuclei and Stars}'' and a research grant from the Province of Styria. G.~Endr\H{o}di acknowledges support from the EU (ITN STRONGnet 238353).

\end{document}